# Quantum Chaos in Open versus Closed Quantum Dots: Signatures of Interacting Particles

C. M. MARCUS, S. R. PATEL, A. G. HUIBERS, S. M. CRONENWETT, M. SWITKES, I. H. CHAN,
R. M. CLARKE, J. A. FOLK, S. F. GODIJN
*Department of Physics, Stanford University, Stanford CA 94305-4060 U.S.A.*

K. CAMPMAN AND A. C. GOSSARD
*Materials Department, University of California at Santa Barbara, Santa Barbara CA 93106 U.S.A.*

*This paper reviews recent studies of mesoscopic fluctuations in transport through ballistic quantum dots, emphasizing differences between conduction through open dots and tunneling through nearly isolated dots. Both the open dots and the tunnel-contacted dots show random, repeatable conductance fluctuations with universal statistical properties that are accurately characterized by a variety of theoretical models including random matrix theory, semiclassical methods and nonlinear sigma model calculations. We apply these results in open dots to extract the dephasing rate of electrons within the dot. In the tunneling regime, electron interaction dominates transport since the tunneling of a single electron onto a small dot may be sufficiently energetically costly (due to the small capacitance) that conduction is suppressed altogether. How interactions combine with quantum interference are best seen in this regime.*

## 1. Overview of Ballistic Conductance Fluctuations from Open to Isolated Quantum Dots

Electron transport through quantum dots, i.e. micron-scale confined conductors, is an exceedingly rich experimental system, bearing signatures of quantum interference, discreteness of charge, electron-electron interaction, chaotic dynamics of quasiparticles, and decoherence. This paper describes recent experiments in which interference and interactions both play important roles—interference as the source of random mesoscopic conductance fluctuations, and interactions causing quasiparticle dephasing as well as to the so-called Coulomb blockade of conduction at low bias. Our motivation is to begin to address the difficult problem of how strong electron-electron interactions affect quantum interference.

The paper is divided into five sections. In this section, an overview and short history of conductance fluctuations in open ballistic quantum dots and Coulomb blockade in tunneling dots is given, with some typical experimental examples. In Sec. 2, basic fabrication and experimental methods are discussed. In Sec. 3 we show how to use the universality of mesoscopic effects in open dots to measure the dephasing rate using the change in average conductance upon breaking time reversal symmetry (weak localization). In Sec. 4 we consider nearly isolated quantum dots and demonstrate the universal distribution of Coulomb blockade peak heights and the parametric dependence of Coulomb peaks on magnetic field. In Sec. 5, we consider mesoscopic fluctuations of cotunneling in the valleys between Coulomb peaks and show that elastic cotunneling exhibits random magnetoconductance fluctuations with a larger characteristic field than that of the peaks, an effect which can be interpreted in terms of short the tunneling time $\sim \hbar/E_c$ where $E_c = e^2/C$ is the charging energy, associated with this conduction process. Sec. 6 contains concluding remarks. Parts of this text are modified from concurrently appearing conference proceedings.

In disordered metallic or semiconductor samples with large conductances, $G \gg e^2/h$, electron-electron interactions lead to small corrections to transport and contribute to quasiparticle dephasing [1]. When such systems are comparable in size to the dephasing length $\ell_\varphi$, mesoscopic fluctuations in sample conductance have universal statistical properties which do not depend on material properties or average conductance of the sample, but only on a few basic symmetries of the system. For this reason they are called universal conductance fluctuations (UCF) [2-4]. In the past few years, experiments using high-mobility semiconductor quantum dots have shown that confined ballistic structures (i.e. devices in which the bulk elastic mean free path $\ell$ as well as the dephasing length $\ell_\varphi$ exceeds the sample dimensions) also exhibit mesoscopic conductance fluctuations similar to disordered systems [5-9] with the same universal statistics [10, 11]. The basic phenomenon of universal conductance fluctuations in a ballistic structure is shown in Fig. 1 for a relatively large quantum dot of area $\sim 2$ $\mu m^2$ fabricated from a GaAs/AlGaAs heterostructure.

Random conductance fluctuations are observed at low temperature as a function of an external magnetic field applied perpendicular to the plane of the dot. The fluctuations are repeatable within a single cooldown of the device and, as can be seen in Fig. 1, are symmetric in B, as expected from the Landauer-Büttiker relations. Both the vertical scale of the fluctuations (i.e. the typical fluctuation amplitude $\delta g$) and the horizontal scale (i.e. the characteristic field scale) of the fluctuations are universal, dictated only by the symmetries of the system and simple scaling parameters, respectively. Such measurements and related theory have drawn attention to the fact that *disorder is not a requirement for UCF*, but is only one means of generating the universal features of quantum transport. It is now widely appreciated that universalities of UCF are intimately related to the universal statistics of quantum systems whose classical analogs are chaotic [12-14], irrespective of disorder. Universal statistics in ballistic microstructures are therefore applicable whenever, but only when, the devices have irregular or "chaos-generating" shapes. (In classical dynamics the generic situation is a mixed phase space, with some trajectories executing regular motion and others chaotic motion. A mixed phase space can lead to fractal conductance fluctuations [15, 16]).

Following the suggestion by Jalabert, Baranger and Stone [17] that open ballistic quantum dots will show UCF and so provide a useful test system for quantum manifestations of classical chaos, a large theoretical effort has succeeded in connecting UCF and other mesoscopic effects to universal spectral properties of random matrices, nonlinear sigma models, semiclassical techniques, and quantum manifestations of classical chaos [14, 18]. "Open" quantum dots means that the dot is connected to the bulk electron gas via leads that carry several quantum modes, $N_i \gg 1$ for the $i^{th}$ lead. Typically, conductance out of the dot, $g_{tot} = (2e^2/h)\sum_i N_i$, is much greater than $e^2/h$. In the open regime, lifetime broadening $\Gamma \sim h/\tau_{escape}$ of quasi-bound states of the dot exceeds the mean level spacing $\Delta = 2\pi\hbar^2/m^* A_{dot}$ (= $7.1\mu V/A_{dot}[\mu m^2]$ = 83 mK/$A_{dot}$ [$\mu m^2$] in GaAs), and, for dots much larger than the Fermi wavelength $\lambda_F$ (~40nm in GaAs/AlGaAs heterostructures), a semiclassical picture of interfering classical trajectories bouncing within the dot is appropriate. If, in addition, the motion of electrons in the dot is classically chaotic (due either to disorder or the shape of the confining potential) a random scattering matrix approach [19] or nonlinear sigma model [20] is appropriate for calculating transport properties. These methods are consistent with one another, and agree quantitatively with experiments in open dots (e.g. [9, 21]) once finite temperature and dephasing are accounted for [20-23]. Note that these theoretical approaches describe single-particle physics with dephasing added as a phenomenological parameter. Given the good agreement with experiment, this is apparently a reasonable approximation for describing transport through open quantum dots.

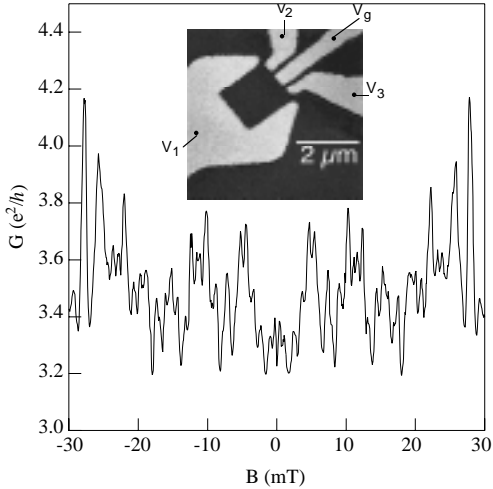

**Fig. 1**. Typical conductance fluctuations of a ballistic quantum dot at base temperature of the dilution refrigerator, with electron temperature $T_e \sim 100$ mK. The average conductance indicates that $\sim 3$–$4$ channels in each lead are open. Note symmetry in field, and typical field scale of order one flux quantum through the dot. Inset. The gate labeled $V_g$ is used to induce shape fluctuations, while $V_1$ and $V_2$ set the point contact conductance.

Theories of conductance fluctuations generally address statistical features of transport: averages, variances, distributions, correlations. To compare with experiment, the system being measured must be able to generate statistics. The variance of UCF can be studied by sweeping the magnetic field B if one assumes that long sweeps of magneto-conductance fluctuations constitute a stationary distribution, that is, that sweeping B *only* generates UCF and the distribution itself does not depend on B. This is almost never true. For instance in ballistic dots as the cyclotron radius becomes comparable to the mean free path, or the size of the device in the ballistic case, the dynamics crosses over from chaotic to regular. Moreover using B to study statistics one cannot investigate statistics at some fixed B, say B=0. To get around this problem, and to allow very large statistical samples (which are necessary to find full distributions as opposed to moments) we have developed the technique of shape distortion as a means of generating UCF, following a suggestion by Bruus and Stone [24]. In Fig. 1, the voltage on the center "pin" gate between the point contacts can be swept, putting a continuously adjustable "dent" in the dot. UCF in the 2D landscape of field B and shape distortion, (controlled by $V_g$) is shown in Fig. 2. Figure 2 also shows the full distribution of conductance fluctuations, after subtracting off the local average conductance. The distribution is apparently well described by a gaussian, which is what one expects for a metallic sample (with many conducting modes per lead) but *not* for a few mode quantum dot at T = 0. However, the inclusion of dephasing and thermal smearing into an the random matrix model indeed predicts a nearly gaussian distribution in the dot case as well. (Although deviations from a gaussian in the limit of large dephasing persist for single-mode leads [23].)

For single-pin dots like the one shown in Fig. 1, shape distortion that arises from a changing pin-gate voltage has two contributions, area distortion and pure shape distortion, each having different correlation properties. Area distortion changes the level spacing and so is comparable to energy-induced fluctuations, while pure shape distortion constitutes a true parametric fluctuation like changing magnetic field. The data in Fig. 2 is a mixed case. In dots with multiple pins (described below) the two contributions can be separated, by either simultaneously "pushing" on both pins (electrostatically) for area distortion, or pushing on one and pulling on the other for pure shape distortion.

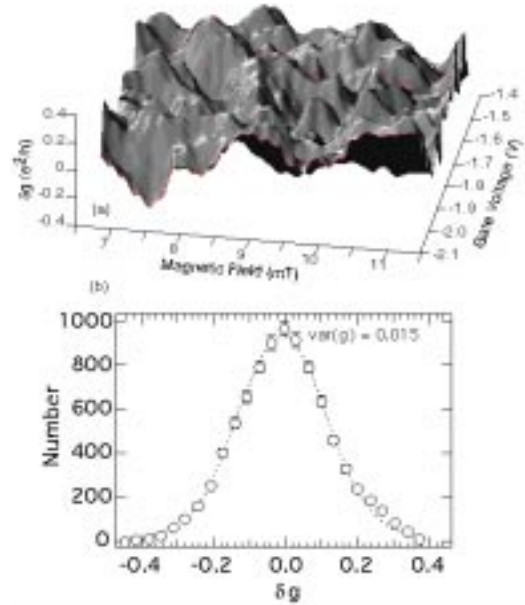

**Fig. 2**. (a) Random landscape of conductance as a function of magnetic field B and shape distortion controlled by center gate voltage labeled $V_g$ in Fig. 1. (b) The distribution of conductance fluctuations P($\delta$g) (open circles) is approximately gaussian (dashed curve). The gaussian distribution as well as the value of var(g) indicate that several channels of phase breaking are present, $N_\varphi$ ~8±3. Modified from Ref. [9].

As point contact leads are pinched off electrostatically and the conductance of each lead falls below $2e^2/h$, transport through the dot occurs via tunneling only. In the tunneling regime, low-bias conductance is suppressed due to classical charging effects (the Coulomb blockade) whenever the energy to add a single electron to the quantum dot, $e^2/C$, where C is the total capacitance of the dot, exceeds the source-drain voltage $V_{sd}$ [25, 26]. When the dot potential is tuned via electrostatic gates so that the number of electrons can fluctuate by one without energy cost, the Coulomb blockade is lifted and a large conductance peak appears at this value of gate voltage, as illustrated in . Because the Coulomb blockade is a classical effect, it appears over a broad temperature range, $kT < e^2/C$ (up to ~4 K for micron-scale GaAs dots).

At lower temperatures, $kT <~\Delta$, quantum-confined levels in the dot become resolved and the Coulomb blockade peaks begin to grow with decreasing temperature (as seen in Fig. 3). This is also the regime where Coulomb peaks show mesoscopic fluctuations in height, both as a function of peak index and magnetic field. Physically Coulomb peak height fluctuations arise from fluctuations in the spatial pattern of the wave function in the dot which changes the coupling of the dot to the leads as external parameters are swept. Random matrix theory predictions for the statistics of these fluctuations in the regime $\Gamma << kT << \Delta$, were made by Jalabert, Stone and Alhassid [27] and were recently verified experimentally by Chang *et al.* [28] and by our group [29].

Between Coulomb blockade peaks, resonant conductance is exponentially suppressed but conductance remain finite due to higher-order elastic and inelastic transport processes collectively referred to as cotunneling [30, 31]. At low temperature and source-drain voltage, $(kT, eV_{sd}) < (\Delta e^2/C)^{1/2}$, cotunneling is dominated by an elastic mechanism in which an electron (or hole) passes virtually through the charged state of the dot with energy of order $\sim e^2/2C$ above the Fermi energy. This process is classically forbidden but can take place if the time is sufficiently short, $\tau \sim \hbar/(e^2/C)$ as dictated by the uncertainty relation. The short time scale for the virtual process shows up experimentally as an increase in the magnetic field correlation length of the magneto-cotunneling fluctuations compared to the on-peak fluctuations (where the characteristic time is set by the inverse level spacing). A recent theory of cotunneling fluctuations by Aleiner and Glazman [32] quantify these relations

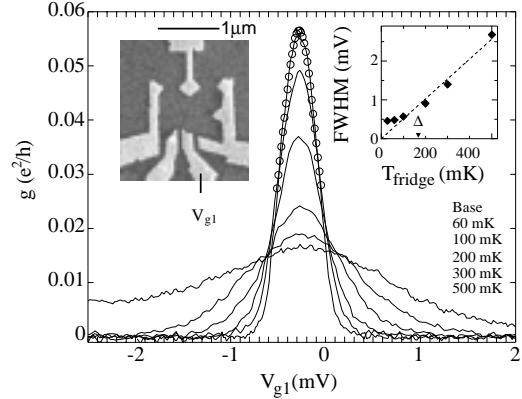

**Fig. 3.** (a) Coulomb blockade peak in conductance g as a function of gate voltage $V_{g1}$ for several temperatures for a quantum dot with level spacing $\Delta = 15\mu eV$. Circles show fit of base temperature peak to $g/g_{max} = \cosh^{-2}(\eta eV_g/2kT)$ Left inset: Micrograph of the quantum dot. $V_{g1}$ is one of two shape-distorting gates. Right Inset: Peak width measured as FWHM from the fit to $\cosh^{-2}$. Linear behavior at high temperatures gives voltage-to-energy scale $\eta=0.12$ for this dot based on a fit to the high-temperature ($kT > \Delta$) result, FWHM ~ 4.3 kT. Saturation at low T is presumably due to electron heating above the refrigerator temperature. Modified from Ref. [29]

as well as give the full statistical distribution of co-tunneling fluctuations.

The appearance of the Coulomb blockade as the quantum dots becomes nearly isolated from the bulk reservoirs is the first and simplest many-body effect to dramatically alter transport. For this reason, the transition between the noninteracting quantum interference of Fig. 1 to the strong influence of electron-electron charging effects in Fig. 3 is an important path, and should be followed carefully.

**2. Experimental Aspects**

The quantum dots discussed in this paper were fabricated on Si-delta-doped GaAs/AlGaAs heterostructure material with a two dimensional electron gas (2DEG) 50 – 160 nm below the surface, depending on the sample used. The dots were formed by electrostatic depletion using Cr/Au gates defined by electron beam lithography. Typical mobilities of the bulk electron gas are in the range 0.5–1.5 × $10^6$ cm$^2$/Vs and bulk density n ~ 3 × $10^{11}$ cm$^{-2}$ give a Fermi wavelength $\lambda_F$ ~40 nm and a transport mean free path $\ell$ = 9 µm, much greater than the 0.5–2 µm lateral dimension of the dot, so that transport within the dot is ballistic. The dots are coupled to bulk 2DEG by individually adjustable point contact leads. Other gates, isolated by gaps from the point contacts, allow small changes to shape and area without affecting point contact conductance, except for a small stray capacitive coupling. When the dot is properly depleted, current cannot escape through these gaps. The dots are irregularly shaped and expected to be chaotic at low magnetic fields, i.e. as long as the cyclotron radius $r_{cyc}$ = h/$\lambda_F$eB ~ 90 nm/B[T] is much less than the lateral size of the dot [33]. Straight-through and directly reflected trajectories are suppressed by the dot shape and the orientation of the point contact leads. Such nonergodic trajectories can dramatically affect the fluctuation statistics in open dots [34], and even affect such properties as the dependence of fluctuation amplitude on temperature [35]. Geometrical effects associated with nonchaotic trajectories in the Coulomb blockade regime have not been investigated.

However, when a quantizing magnetic field ($r_{cyc}$ << L) is applied to a Coulomb blockaded dot, periodic modulations of peak heights and positions, as studied in detail by McEuen and coworkers [36-38] and others more recently, for instance Heinzel *et al.* [39]. Conductance was measured in a dilution refrigerator with a base temperature of $T_{base}$ = 30 mK and a $^3$He cryostat which operates over a continuous temperature range 0.39K < T < 100K. The temperature of the electrons in the dilution refrigerator appears limited to $T_e$ > ~100 mK based on the saturation of Coulomb peak height (Fig. 3 inset) and amplitude of conductance fluctuations (Fig. 3 inset of Ref. [40]).

In the open regime, dot conductances are measured using a standard 4-wire current bias technique through four NiAuGe ohmic contacts using a PAR 124 lock-in amplifier, at typically <100 Hz. Gates are referred to lock-in low using batteries and computer controlled D/As run through instrumentation amplifiers. In the Coulomb blockade regime, the dot resistance is typically much higher than lead wire resistance, and is measured using a 2-wire voltage bias with an Ithaco 1211 current preamplifier at the front end of a PAR 124 lock-in amplifier at 11 Hz. The gates are referred to a free ohmic contact on the Ithaco-low side of the dot through a 1GΩ resistor. In either measurement configuration, the total voltage dropped across the dot is kept below 5µV$_{rms}$. All leads are rf filtered using Spectrum Control multipin Pi filters at room temperature. No copper powder or other cryogenic filtering is used, although cold 1kΩ series resistors are used on the dilution refrigerator, which is also in an rf shielded room.

**3. Open Dots: The Effects of Dephasing**

The statistical properties of conductance, G, in open quantum dots —not just the mean, ⟨G⟩ and variance, var(G), but the full distribution P(G)—is predicted to be universal at T = 0 for dots with chaotic classical dynamics, depending only on the number of quantum modes, N, entering and leaving the dot. However, the "universality" in UCF is even stronger: for large N (in practice N >~3 – 4 suffices), *only* $\langle G \rangle$ depends on N, since P(G) becomes gaussian and var(G) approaches a *constant*, depending only on whether time reversal symmetry is obeyed (B=0) or broken (B≠0)

$$\text{var}(G) \xrightarrow{N \to \infty} \left(\frac{2e^2}{h}\right)^2 \times \begin{cases} 1/8 & (B = 0) \\ 1/16 & (B \neq 0) \end{cases} \quad (1)$$

where the factor of 2 explicitly accounts for spin. Also at large N, the difference between $\langle G \rangle_{B \neq 0}$ and $\langle G \rangle_{B=0}$, the so-called weak localization correction, approaches a constant,

$$\begin{aligned} \delta \langle G \rangle &\equiv \left(\langle G \rangle_{B \neq 0} - \langle G \rangle_{B=0}\right) \\ &\xrightarrow{N \to \infty} 1/4 \left(2e^2/h\right). \end{aligned} \quad (2)$$

At any finite number of channels per lead, exact expressions for $\langle G \rangle$ and var(G) giving the dependence on N are known within RMT [10, 11].

Finite temperature alters these universal results, reducing both the amplitude of the fluctuations, Eq. (1), and the difference in averages, Eq. (2). Fluctuation amplitude is reduced by two mechanism, thermal smearing, resulting from a superposition of ~ kT/Γ independent fluctuation patterns, and dephasing of quasiparticles, primarily due to electron-electron scattering at these temperatures. In contrast, thermal averaging does not affect $\delta \langle G \rangle$, so its reduction with higher temperature is purely the result of dephasing. Physically, the dephasing time $\tau_\varphi$ represents the quantum coherent lifetime of a quasiparticle during which it may self-interfere. It is finite at nonzero temperatures because of the interaction of the quasiparticle with the environment (including other quasiparticles) which effectively measures the position of the quasiparticle and thus eliminates interference [41]. Because dephasing is an exponential process (i.e. equal probability per unit time) similar to escape from a chaotic system, it can be accounted for within RMT as an effective voltage probe that carries $N_\varphi$ channels but allows no net particle flux. Including these channels into a RMT model of conductance [21, 22] an approximate expression for $\delta \langle G \rangle$ in terms of the dephasing rate $\gamma_\varphi = 1/\tau_\varphi$ [21],

$$\delta \langle G \rangle \cong \frac{N}{2N + N_\varphi}\left(e^2/h\right) \quad (3)$$

where $N_\varphi$ is the number of dephasing channels,

$$\begin{aligned} N_\varphi &\equiv 2\pi\left(\frac{\hbar \gamma_\varphi}{\Delta}\right) = \\ &0.58 \, A_{dot}\left[\mu m^2\right] \gamma_\varphi \left[ns^{-1}\right]. \end{aligned} \quad (4)$$

Aleiner and Larkin showed that Eq. 3, originally constructed by Baranger and Mello [21] as an interpolation between two known results, is exact for large N, $N_\varphi \gg 1$, and that N and $N_\varphi$ need not be integer valued (as needed in a random matrix model) [42]. Very recently, Brouwer and Beenakker [23] presented a more complete picture of dephasing in quantum dots, unifying the voltage-probe model with the imaginary-potential model used for instance in Ref. [43] and again obtained Eq. 3 in the limit (N, $N_\varphi$) >>1, but find significant corrections for N=1, including a deviation from gaussian statistics in the limit $N_\varphi \to \infty$. Already for N = 2, however, their results are quite close to Eq. 3.

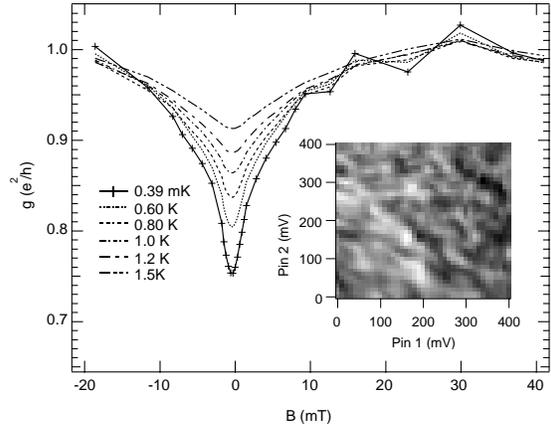

**Fig. 4.** Shape-averaged conductance g(B)=$\langle G(B)\rangle_{shape}$/ ($e^2/h$) N=1 channel per lead for the dot pictured in the right inset of Fig. 5. The reduction in conductance at B=0, known as weak localization or quantum enhanced backscattering, results from quantum interference of time reversed trajectories, or, equivalently, from time reversal symmetry of the scattering matrix. The reduction is eliminated upon breaking time reversal symmetry with a magnetic field comparable to ~1 flux quantum through the dot. The size of the conductance dip $\delta \langle G \rangle$ gives a measure of the dephasing rate according to Eqs. 3 and 4. Inset: Raster over shape distorting pins showing shape-induced fluctuations. For each point in the curves g(B), 16 points covering pin-pin space were averaged [46].

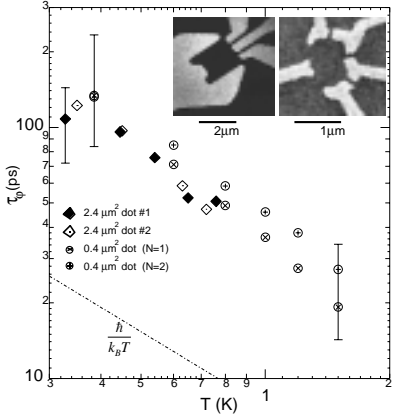

**Fig. 5.** Phase coherence time $\tau_\varphi(T)$ for two different sized dots, measured using different methods. The small dot data (circles) were extracted from the weak localization dip in g(B) (see Fig. 4) using Eqs. 3 and 4 for N=1 and N=2. The large dot data (diamonds) were extracted using a semiclassical method described in the text [40]. $\tau_\varphi(T)$ in these open dots appears consistent with theory for 2D electron-electron scattering theory.

Using Eq. 3 and its refinement in Ref. [23], we can now use the weak localization amplitude $\delta\langle G\rangle$ to find the temperature dependence of the dephasing rate, $\tau_\varphi(T)$. The form of $\tau_\varphi(T)$ in zero dimensions (i.e. in a dot) is a controversial theoretical problem and is some technological significance, particularly given recent discussions of quantum computation, and its limitations due to decoherence. The shape-averaged conductance $g(B) = \langle G(B)\rangle_{shape}/(e^2/h)$ is shown in Fig. 4 for temperatures between 0.39K and 1.5K, for a 0.4 $\mu m^2$ quantum dot at N=1 modes in each lead. (A micrograph of the device is shown as the right inset in Fig. 5.) The inset of Fig. 4 shows the two-dimensional shape-distortion landscape over which averaging was carried out. Each point in the various g(B) traces represents an average over 16 samples covering the pin voltage range shown, 0 – 400 mV on each pin. The extracted values of $\tau_\varphi(T)$ are shown in Fig. 5.

Also shown in Fig. 5 are data from a considerably larger dot extracted using a semiclassical method developed in Refs. [40, 44, 45]. This method uses changes in the characteristic magnetic field scale $B_c$ of conductance fluctuations as the leads are open. The argument [40] is again based on an effective-lead approach, and takes advantage of the fact that the effective escape rate, the sum of escape and dephasing rates, is proportional to $(B_c)^{-1/2}$ due to the diffusive accumulation of area by a chaotic trajectory. The value of $B_c$ extrapolated to the limit of closed leads therefore corresponds pure due dephasing, with the constant of proportionality given by the slope of real escape (which is proportional to $\langle G\rangle$) versus $(B_c)^{-1/2}$. The consistent values for $\tau_\varphi(T)$ found using these different methods, measured on dots whose area differed by a factor of six, makes one confident in the validity of these approaches and of the data in Fig. 5.

At 400 mK, the ballistic dephasing length $\ell_\varphi = v_F \tau_\varphi$ is ~25 $\mu$m, a few times the bulk mean free path. The data appear consistent with 2D dephasing (theory and experiment) and appears well represented by a mixture of dephasing mechanisms with differing exponents $\tau_\varphi^{-1}(T) \sim AT + BT^2$. Moreover, the values found are within a factor of two of the 2D electron-electron scattering theory of Altshuler and Aronov [1] if one naively replaces the mean free path in the theory with the lateral dimension of the dot. Details and further measurements of dephasing in quantum dots as a function of temperature and applied current bias will appear in a subsequent publication [46].

## 4. Isolated Dots: Fluctuations of Coulomb Blockade Peaks

As the gate voltages controlling the sizes of the point contact leads are made more negative, lead conductances will eventually fall below the single-channel conductance, $G < 2e^2/h$, and act as tunnel barriers. The sharp edge where each lead conductance shuts off is shown in Fig. 6. This 2D raster over point contact gate voltages also indicates how experimentally the two point contacts can be balanced for arbitrary dot conductance.

At low temperatures and voltages across the dot, $(eV_{sd}, kT) < \sim\Delta$, the Coulomb blockade peaks (as a function of gate voltage) show large height fluctuations—sufficient to cause some peaks to disappear entirely—as seen in Fig. 7(a). The on-peak conductance is mediated by resonant tunneling through one or a few eigenstates of the dot. The fluctuations in height result from changes in the coupling of the dot wave function(s) to the bulk electron states just outside the leads. Physically, it is the projection of the "speckle pattern" of the dot wave function onto the plane-wave state in the lead that determines its coupling. The statistics of wave functions therefore control the statistics of peak heights. Jalabert, Stone and Alhassid [27] used random matrix theory (RMT) to derive universal peak height distributions both for systems with time-reversal symmetry (orthogonal ensemble, B=0) and without (unitary ensemble, B≠0). Their results have subsequently been extended to nonequivalent [43] and multimode [47] leads and to systems whose dynamics cross the transition to from regular to chaotic [24].

These studies consider a disordered or chaotic-ballistic quantum dot coupled to the two reservoirs (labeled 1 and 2) by tunneling leads with tunneling rates $\gamma_{1(2)} = \Gamma_{1(2)}/\hbar$. In the regime $\Gamma \ll kT \ll \Delta$ (where $\Gamma = \Gamma_1 + \Gamma_2$) blockade peaks have roughly equal width $\sim kT$, and height $g_{max}$ that depends on the coupling of the leads to the dot [27],

$$g_{max} = \frac{e^2}{h}\frac{\pi}{2kT}\frac{\Gamma_1\Gamma_2}{\Gamma_1+\Gamma_2} \equiv \frac{e^2}{h}\frac{\pi\overline{\Gamma}}{2kT}\alpha \qquad (5)$$

where $\alpha \equiv \Gamma_1\Gamma_2/\overline{\Gamma}(\Gamma_1+\Gamma_2)$ is a dimensionless parameter that contains all of the fluctuations in $g_{max}$ arising from changes in $\Gamma_1$ and $\Gamma_2$. If the leads are balanced (using a raster as shown in Fig. 6) and located far away (many $\lambda_F$'s from one another, we may take the leads to be *statistically* identical and independent ($\overline{\Gamma}_1 = \overline{\Gamma}_2 = \overline{\Gamma}/2$). In this case, RMT predicts a universal peak height probability distributions for zero and nonzero magnetic fields,

$$P_{(B=0)}(\alpha) = \sqrt{2/\pi\alpha}\, e^{-2\alpha} \qquad (6a)$$

$$P_{(B\neq 0)}(\alpha) = 4\alpha(K_0(2\alpha) + K_1(2\alpha))\, e^{-2\alpha} \qquad (6b)$$

where $K_0$ and $K_1$ are modified Bessel functions [27, 43]. Average peak heights in zero and nonzero field are given by inserting the averages of $\alpha$,

$$\int \alpha\, P_{B=0}(\alpha)d\alpha = 1/4, \quad \int \alpha\, P_{B\neq 0}(\alpha)d\alpha = 1/3,$$

into Eq. (1). Peak height distributions in Fig. 7 (a) and (b) were measured by sweeping one shape-distorting gate voltage $V_{g1}$ repeatedly over one or several peaks then changing either the magnetic field or the other gate voltage. $V_{g1}$ can be swept over ~40 peaks before the average peak height begin to change. Using the second pin allows a much larger data set to be gathered at a single value of magnetic field. Additionally, for the B≠0 data, several values of magnetic field over the range ~50 - 150 mT were used to provide more data. For each ensemble, statistics from a number of data sets were combined after first transforming each data set to the scaled variable α using a two-parameter fit to Eq. 6. Finally, histograms of the combined data were normalized to unity area. By plotting P(α) on a logarithmic scale (unlike in original published version, Ref. [29]) the agreement between theory and experiment is seen to extend out to very large peak heights [48]. As seen in the inset of (a), neighboring peak heights are correlated over ~4–5 peak spacings at base temperature, with increasing correlation at higher temperature. Each distribution in Fig. 7 represents ~600 peaks, but because of this correlation, we estimate that each figure consists of roughly ~90 statistically independent peaks.

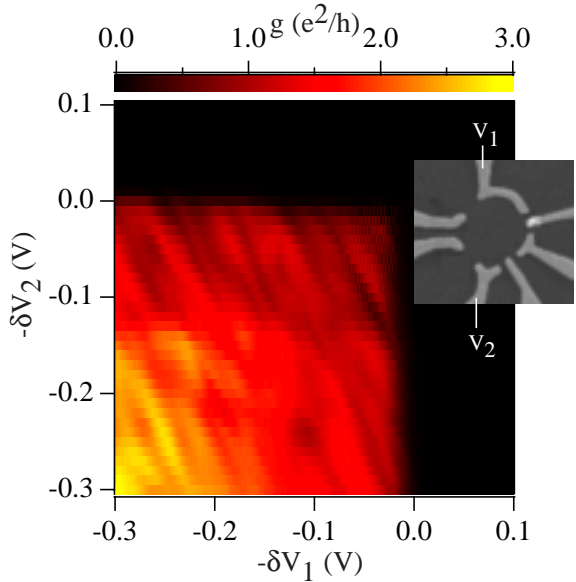

**Fig. 6.** Color-scale plot of conductance as a function of the gate voltages controlling the point contacts. The abrupt shut-off in conductance near $\delta V_{1,2} = 0$ allows the point contacts to be balanced. Coulomb blockade peaks are strongest near the corner, $\delta V_{1,2} = 0$.

The fact that neighboring peak heights are correlated is unexpected within a single-particle-plus-charging-energy picture. Since temperature contributes to this correlation, one might suppose that having $T \sim \Delta/2$ rather than $T \ll \Delta$ may be the cause of the correlation. However, this appears to be in conflict with the good agreement with the theoretical distributions seen in Fig. 7. These distributions also have a strong temperature dependence [49]. Indeed, similar experiments with smaller dots ($\Delta \sim$ 600 mK and 50-100 electrons in the dot) by Chang *et al.* may show less correlation between neighboring peaks [28]. Any remaining correlation in peak heights in the limit $T \ll \Delta$, would be beyond a single-particle RMT, and raises the interesting prospect that the single-particle RMT approach reproduces the experimentally observed distributions (as seen in Fig. 3, and in Ref. [28]) but fails to predict peak-to-peak correlations which may arise either due to interactions or nonuniversal wave function statistics. This issue is the subject of experiments currently underway.

Two-dimensional raster-sweeps over $V_{g1}$ and B, as shown in Fig. 8, allow fluctuations in peak height and position to be measured as a quasi-continuous function of an external parameter. Peak height as a function of B can be obtained in post-processing of data by following maxima in the $V_{g1}$–B plane. The resulting parametric peak heights $g_{max}(B)$ are insensitive to small jumps in peak position due to switching noise. Like UCF measurements of G(B) in open dots, $g_{max}(B)$ appears random and symmetric about B=0 and repeatable within a cooldown of the device. Examples of $g_{max}(B)$ are shown in Fig. 9 (a) and (c).

An external magnetic field couples to the electrons in the dot as an Aharonov-Bohm flux which scrambles the speckle pattern of the wave function in the dot. This in turn scrambles the tunneling rates $\Gamma_{1(2)}(\varphi)$ through the leads and hence (by Eq. (5)) the peak height [50, 51]. How much flux is needed to scramble the wave function? Defining a dimensionless flux $\varphi = BA_{dot}/\phi_o$, we consider the autocorrelation of peak height fluctuation as a function of flux,

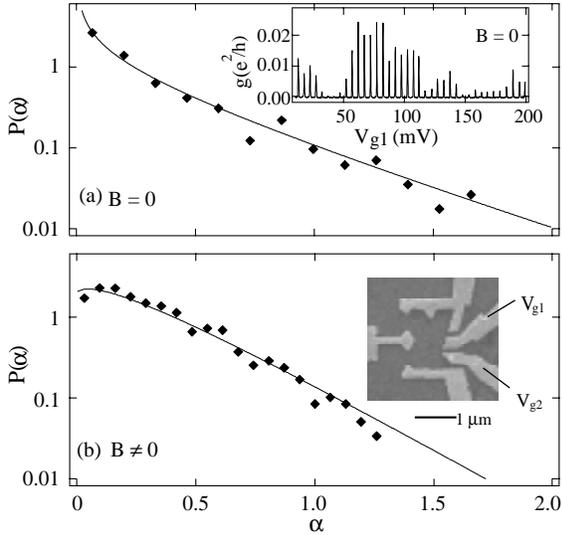

**Fig. 7.** Distributions of peak heights in dimensionless units $\alpha \equiv \Gamma_1\Gamma_2/((\Gamma_1+\Gamma_2)\Gamma)$ (see Eq. 5) for B=0 (a) and B≠0 (b). Solid curves are random matrix theory predictions, Eq. 6. Top Inset: Sample of peaks used to produce B=0 distribution. Bottom Inset: Electron micrograph of the device measured to produce the peak height distributions. Labels $V_{g1}$ and $V_{g2}$ indicate shape-distorting gates.

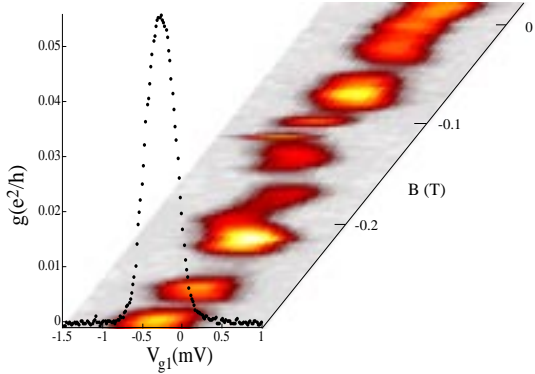

**Fig. 8.** Color-scale plot of parametric fluctuations of height and position of a single Coulomb blockade peak as a function of B. The rms excursion of the peak position, scaled from gate voltage to energy (using peak width vs. T in Fig. 3 ) is ~0.55$\Delta$.

$$C_{peak}(\Delta\varphi) = \left\langle \tilde{g}_{max}(\varphi)\, \tilde{g}_{max}(\varphi + \Delta\varphi) \right\rangle_\varphi / \text{var}(g_{max})_\varphi,$$

where $\tilde{g}_{max} = (g_{max} - \langle g_{max} \rangle)$ is the deviation of the peak height from its average. This function has been calculated for $\Gamma \ll kT \ll \Delta$ within a noninteracting RMT approach perturbatively near $\Delta\varphi = 0$ and numerically over the full range of $\Delta\varphi$ [50, 51]. In the unitary (B≠0) ensemble, the numerically obtained universal function $C_{peak}(\Delta\varphi)$ can be approximated as Lorentzian squared

$$C_{peak}(\Delta\varphi) \sim \frac{1}{\left(1 + \left(\Delta\varphi / 0.54\, \varphi_c^{peak}\right)^2\right)^2} \quad (7)$$

where the characteristic magnetic flux $\varphi_c^{peak}$ defines a universal scaling parameter, $\sqrt{C_v(0)} = 1/\varphi_c^{peak}$, analogous to the universal correlations of energy level velocity defined by Szafer, Simons and Altshuler [52-54]. Random matrix theory as well as semiclassical arguments give a value for the characteristic flux which is less than one (i.e. less than one flux quantum through the dot) by the ratio,

$$\varphi_c^{peak} \sim \left(\Delta / \kappa E_T\right)^{1/2} \sim \left(2\pi \kappa^2 N_{dot}\right)^{-1/4} \quad (8)$$

where $E_T$ is the Thouless energy, defined for ballistic dots as the inverse time of flight across the dot, $E_T = hv_F / (A_{dot})^{1/2}$, $N_{dot}$ is the number of electrons on the dot, and $\kappa$ is a geometrical factor ($\kappa \ll \sim 1$) [24, 51, 55, 56]. (Equation (8) more or less follows the notation of Bruus *et al.* [51].) Experimentally, we find $C_{peak}(\Delta\varphi)$ is reasonably fit by Eq. 7 with a $\varphi_c^{peak}$ in the range 1.2–2 for several of the devices measured. Numerical estimates for $\kappa$ (as defined in (8) for the case of winding around a flux line) give $\kappa \sim 1/\sqrt{2\pi}$ [51], that is, $\varphi_c^{peak} \sim N_{dot}^{-1/4}$. Therefore the theoretically expected value for the dots in Fig. 9, which have $N_{dot}$ of order 1000, is $\varphi_c^{peak} \sim 1/5$, considerably smaller than the experimental value. It is possible that the difference between flux-winding statistics and areal statistics can help account for this some of this difference; temperature is not expected to change $\varphi_c^{peak}$, by analogy to the situation in open dots [20]. A technique for measuring $\kappa$ directly is given in Ref. [40], but so far has not been carried out for these devices. Parametric correla-tions of density-of-states fluctuations have been measured in vertical transport by Sivan, *et al.* [57]. In their system the characteristic magnetic field was close to the value expected from theory. Their system differed from the present one in several ways: transport was vertical rather than lateral (and therefore only sensitive to fluctuations in density of states, not in the coupling to the leads), the dot was 3D rather than 2D and disordered rather than ballistic. The importance of these experimental differences remains to be sorted out.

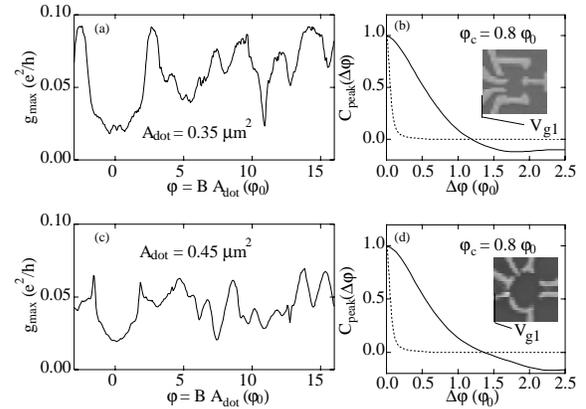

**Fig. 9.** (a,c) Peak height as a function of magnetic field (sweeping over $V_{g1}$) for dots shown in (b,d). Note symmetry in peak height about B = 0. (b,d) Correlation function $C_{peak}(\Delta\varphi)$ averaged over 60 peaks in (b) and 20 peaks in (d) over the magnetic field range 50 – 150 mT (~5 – 15 $\varphi_0$ through the dot) (solid). Theoretical autocorrelation based on semiclassical $\varphi_c^{peak}$ [51]. Note that the measured field scale is larger than semiclassical theory by a factor of ~8 for both devices. Inset: Micrographs of devices used. 2DEG located 800 Å (b), 1600 Å (c) below gates.

Peak position fluctuations are also of interest, and can be related to fluctuations in orbital magnetism [58]. Within a single-particle picture, one expects the amplitude of peak position fluctuations be of order $\Delta$ when scaled to energy units by $\eta$. Using the scaling factor $\eta=0.090$ derived from the line fit in the right inset of Fig. 3 right inset, the rms peak excursion for the sweep in Fig. 8 is 0.55 $\Delta$.

## 5. Fluctuations of Cotunneling Between Coulomb Blockade Peaks

Because low temperature transport on the Coulomb peak is a resonant tunneling process, its mesoscopic fluctuations can be adequately treated a within single-particle picture, as we have seen in the previous section. In this sense, the interactions which give rise to Coulomb blockade in the first place are quenched at the peak (though perhaps are reflected in the large $\varphi_c^{peak}$). Conductance between the peaks is suppressed but is not zero, and it is here that the influence of interactions most clearly coexists with quantum interference.

At temperatures and source-drain voltages satisfying the modest requirement $(kT, eV_{sd}) < (\Delta e^2/C)^{1/2}$, the residual conductance between Coulomb blockade peaks is dominated by *elastic cotunneling* in which an electron (or hole) virtually tunnels through an energetically forbidden charge state of the dot lying at an energy E above (below) the Fermi energy in the leads, where E equals $e^2/2C$ at the center of each valley between peaks and decreases to zero on the peak [59]. (Note that the charging energy is typically much larger than the level spacing. For our dots, $e^2/C \sim 0.3$–0.8 meV while $\Delta \sim 20$ $\mu$eV, thus the allowed range of T and $V_{sd}$ is larger for elastic cotunneling than for resonant tunneling.) Elastic cotunneling is a classically forbidden *virtual* process. It can occur as long as it takes place in a time $\sim\hbar/E$ consistent with the uncertainty relation. Average transport properties for elastic as well as inelastic cotunneling were given by Averin and Nazarov [30]. Aleiner and Glazman recently extended this work to include mesoscopic fluctuations of elastic cotunneling [32].

In strongly-blockaded dots, cotunneling currents are usually very small and so are difficult to measure. However, once the tunneling point contacts are sufficiently open, say $G_{1,2} > \sim 0.5(2e^2/h)$, fluctuations in the valleys can be measured quite easily. A color-scale plot of two peaks and their neighboring valleys is shown in Fig. 10, along with traces showing the location in gate voltage of $g_{max}$ (the peak) and $g_{min}$ (the valley). By measuring conductance long these extreme paths, we can compare fluctuations of peaks and valleys, $g_{max}(B)$ and $g_{min}(B)$, the former well described by single-particle physics, the latter not. Figures 11a and 11b shows $g_{max}(B)$ and $g_{min}(B)$, along the paths indicated in Fig. 10. The autocorrelation $C_{peak(valley)}(\Delta B)$ of $\tilde{g}_{max(min)}(B)$, defined in analogy to $C_{peak}(\Delta\varphi)$ in Sec. 4 is shown in Fig. 11c and illustrates the salient difference between resonant (peak) and cotunneling (valley) fluctuations in the same dot: the characteristic magnetic field $B_c$ is larger for the valleys than for the peaks [29]. This difference in $B_c$ can be understood from a semiclassical point of view as follows: On resonance, the characteristic time that an electron spends in the dot, diffusively accumulating Aharonov-Bohm phase, is $\sim h/\Delta$; in cotunneling, the corresponding time is $\sim h/E$, limited by the uncertainty relation. This suggests a characteristic dimensionless flux in the valleys defined in analogy to Eq. (4) for the peaks,

$$\varphi_c^{valley} \sim (E/\kappa E_T)^{1/2} \qquad (9)$$

and that the ratio of characteristic fluxes and fields is related to the ratio of charging energy to level spacing,

$$\frac{\varphi_c^{valley}}{\varphi_c^{peak}} = \frac{B_c^{valley}}{B_c^{peak}} = \sqrt{\frac{E}{\Delta}} \qquad (10)$$

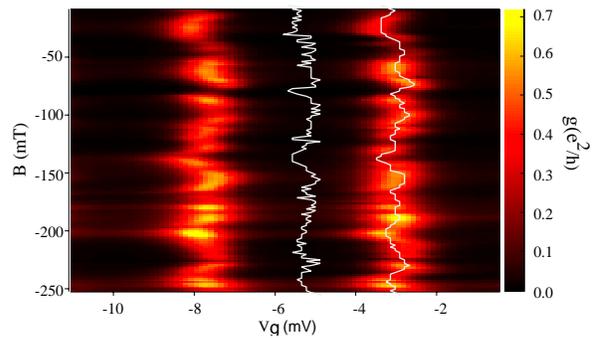

**Fig. 10.** Color-scale plot of conductance (scale indicated on right) showing magnetic field dependence of two subsequent Coulomb blockade peaks. White jagged curves trace $g_{min}(B)$ and $g_{max}(B)$.

Using the experimental parameters, we expect from Eq. 10 a ratio of characteristic fields of $\sim (300\mu V/20\mu V)^{1/2} \sim 4$. This estimate appears inconsistent with the experimentally observed ratio of $\sim 2$ (Fig. 7c). It may be, however, that the source of the disagreement lies more with the peaks than with the valleys, and that on peak some time scale shorter than $h/\Delta$ is acting as the characteristic time for phase accumulation in resonant tunneling.

A detailed theoretical treatment of cotunneling fluctuations [32] accounting for virtual processes through all excited levels above the Coulomb gap reproduces the semiclassical results for the ratios of characteristic fluxes, Eq. (9) and (10), and also predicts a new form for the autocorrelation of valley conductances,

$$C_{valley}(\Delta B) = \Lambda(\Delta B/B_c^{valley}) , \qquad (11a)$$

$$\Lambda(x) = \frac{1}{\pi^2 x^4}\left[\ln x^2 \ln(1+x^4) + \pi \arctan x^2 + \frac{1}{2}Li_2(-x^4)\right]^2. \qquad (11b)$$

We find that Eq. (11) fits the autocorrelation of the valleys better than the squared Lorentzian, Eq. (7), but that for the peaks, Eq. (7) works better, as expected. This will be described in a subsequent publication [60]. The analysis of Ref. [32] goes on to predict full distributions of the cotunneling fluctuations for arbitrary symmetry breaking field. Experimental tests of these distributions are also currently underway [60].

## 6. Conclusions and Open Problems

Mesoscopic fluctuations of conductance through quantum dots bear signatures of not only of the universality of single-particle quantum chaos, but also of the modifications to that universality which result from electron-electron interactions. In open quantum dots, electron-electron interaction is the dominate dephasing mechanism at low temperatures, causing the destruction of quasiparticle quantum coherence. The outstanding challenges here lie in two aspects. First, how does dephasing affect fluctuation statistics? In particular how are the full distributions $P(G)$ and correlations $C(\Delta X)$ with respect to a changes in an external parameter X modified by dephasing? Recent experiments [40, 46, 61, 62] and theory [20-23] have provided many answers here. Second, how does the dephasing rate depend on the dimensionality, size, shape, temperature, disorder, confinement method, and degree of openness of the dot. Here theory remains controversial [63-65] and experiments few [40, 62, 66].

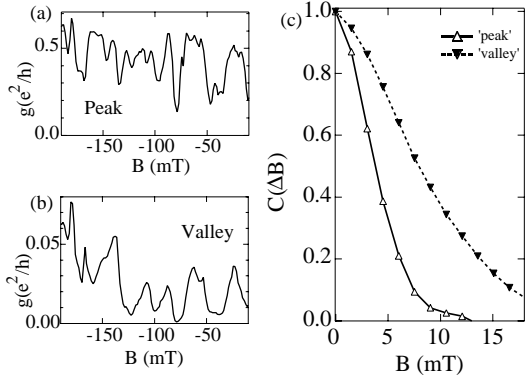

**Fig. 11.** (a) Fluctuations of peak height, $g_{max}(B)$. (b) Fluctuations of elastic cotunneling at $g_{min}(B)$ in adjacent valley. Paths of $g_{min}(B)$ and $g_{max}(B)$ are shown in Fig. 10. (c) Magnetic field autocorrelation $C(\Delta B)$ of the curves in (a) and (b) shown that the valleys have a larger characteristic magnetic field, with $B_c^{valley}/B_c^{peak} \sim 2$. One flux quantum through the dot is $\sim 12$ mT.

In nearly-isolated quantum dot connected to the bulk via tunneling leads, interactions play a more obvious role in transport, suppressing conduction except in narrow peaks as a function of the dot potential (controlled by electrostatic gates). Nonetheless, a single-particle RMT [27] accurately describes the measured fluctuations in peak height in the resonant tunneling regime [28, 29]. An outstanding issue concerns correlations between the heights of neighboring peaks, which are not expected within RMT, but are seen in the experiments on larger dots [29, 48]. The role of finite temperature on fluctuation statistics and correlations between peaks needs to be sorted out. Parametric fluctuations of peaks show general agreement with RMT, but the characteristic field is too large compared to noninteracting RMT results [50, 51, 67]. The role of temperature, dephasing, interactions and uniform field rather than flux have not been addressed. Perhaps one or more of these complications will bring the theory and experiment into accord.

An exciting recent development has been the observation of mesoscopic fluctuations in the valleys between Coulomb peaks [60], and the simultaneous development of a detailed theory of mesoscopic fluctuations of elastic cotunneling [68]. In this problem both quantum coherence and interaction in the form of a charging energy are crucial.

An important issue not discussed in this review but of direct relevance are the mesoscopic fluctuations of peak spacing. Recently, Sivan *et al*. [63] reported that random fluctuation in peak spacing in small gate-defined GaAs quantum dots at low temperature (~100 mK) disagree significantly from the Coulomb-blockade-plus-RMT prediction. They find peak spacing fluctuations up to five times larger than predicted by a naive CB+RMT, with an insensitivity to factor-of-two changes in $\Delta$ as $N_{dot}$ ranges from ~60 to ~120. Also, the distribution of fluctuations does not appear Wigner-Dyson distributed as one might expect for quantum level fluctuations, but is symmetric about its average. These observations have lead Sivan *et al.* to suggest a picture of peak spacing fluctuations that is essentially classical in origin, closely related to the problem of packing charges onto a finite volume. Experiments in larger dots ($N_{dot}$ ~1000) do not show such large fluctuations, which is also the expectation of a recent self-consistent theory. How both temperature and the number of electrons on the dot affect spacing fluctuations is an open problem.

We have also not discussed the beautiful dot-in-a-ring experiments of Yacoby and coworkers [69] which provide a measurement of phase as well as amplitude of the scattering through a disordered dot. The experiment clearly shows a phase shifts of exactly $2\pi$ between neighboring peaks, rather than the expected $\pi$ for a 1D Fabry-Perot type resonant cavity, or a random shift for a 2D chaotic cavity. This behavior continues to defy simple explanation.

Overall, the field of mesoscopic physics is moving into a realm beyond single particle physics where quantum coherence, chaos, confinement, and interactions all play a role. There is no lack of difficult and important problems, experimentally and theoretically.

We thank I. Aleiner, Y. Alhassid, B. Altshuler, H. Baranger, L. Glazman, V. Falko, C. Lewenkopf, E. Mucciolo, and U. Sivan for very stimulating discussions. We grateful acknowledge support at Stanford from the Office of Naval Research under Grant N00014-94-1-0622, the Army Research Office under Grant DAAH04-95-1-0331, the NSF-NYI program, the NSF Center for Materials Research, and the A. P. Sloan Foundation. We also acknowledge support at UCSB by the AFOSR under Grant F49620-94-1-0158, and by QUEST, an NSF Science and Technology Center.